\documentclass{article}
\usepackage[utf8]{inputenc}
\usepackage{url}
\usepackage{authblk}

\title{Three Algorithms for Merging Hierarchical Navigable Small World Graphs}
\author[1,2]{Alexander Ponomarenko \thanks{\texttt{aponom84@gmail.com, aponomarenko@hse.ru}}}
\date{}

\affil[1]{Laboratory LATNA, HSE University}
\affil[2]{Yugabyte, Inc}

\usepackage[numbers,sort&compress]{natbib}

\usepackage{hyperref}

\usepackage{graphicx}
\usepackage{subcaption}
\usepackage{algorithm}
\usepackage{algorithmicx}
\usepackage[noend]{algpseudocode}
\usepackage{amsmath}
\usepackage{amsfonts}
\usepackage{amssymb}
\usepackage{nameref}
\begin{document}

\maketitle

\begin{abstract}
This paper addresses the challenge of merging hierarchical navigable small world (HNSW) graphs, a critical operation for distributed systems, incremental indexing, and database compaction. We propose three algorithms for this task: Naive Graph Merge (NGM), Intra Graph Traversal Merge (IGTM), and Cross Graph Traversal Merge (CGTM). These algorithms differ in their approach to vertex selection and candidate collection during the merge process. We conceptualize graph merging as an iterative process with four key steps: processing vertex selection, candidate collection, neighborhood construction, and information propagation. Our experimental evaluation on the SIFT1M dataset demonstrates that IGTM and CGTM significantly reduce computational costs compared to naive approaches, requiring up to 70\% fewer distance computations while maintaining comparable search accuracy. Surprisingly, IGTM outperforms CGTM in efficiency, contrary to our initial expectations. The proposed algorithms enable efficient consolidation of separately constructed indices, supporting critical operations in modern vector databases and retrieval systems that rely on HNSW for similarity search.
\end{abstract}


\section{Introduction}
\label{sec:intro}
The k-nearest neighbor (k-NN) search problem has emerged as a fundamental computational challenge with critical applications across numerous domains. In today's data-driven landscape, efficient similarity search is essential for recommendation systems, computer vision, multimedia retrieval, and natural language processing applications. Recent advancements in generative AI, particularly Retrieval-Augmented Generation (RAG) systems, have further highlighted the importance of fast and accurate nearest neighbor search. RAG frameworks depend on efficient retrieval of relevant document fragments from large vector databases to ground language model outputs in factual information, making high-performance approximate nearest neighbor (ANN) search algorithms increasingly vital.


\subsection{Nearest Neighbor Search Problem}

Formally, the \(k\)-nearest neighbor search problem can be defined as follows:
Given a set of points
\[
P = \{p_1, p_2, \dots, p_n\}
\]
in a metric space \(\bigl(X, \rho\bigr)\), where \(\rho\) is a distance function, and a query point \(q \in X\), find a set

\[
P_k 
\;=\;
\underset{S \subseteq P,\;|S|=k}{\arg\min}
\;\sum_{p \in S} \rho(q, p)
.
\]
In other words, we seek the set of $k$ points whose total distance to $q$ is minimal.


Three main approaches have been developed for tackling the nearest neighbor search problem:

\begin{enumerate}
    \item \textbf{Space-partitioning trees}: These methods divide the vector space into hierarchical regions, enabling logarithmic search complexity in low dimensions. Notable examples include KD-trees \cite{bentley1990k}, R-trees \cite{guttman1984r}, VP-trees \cite{yianilos1993}, M-trees \cite{ciaccia1997m}, and Cover trees \cite{beygelzimer2006cover}. A comprehensive overview of these approaches for metric spaces is provided by Zezula et al. \cite{zezula2006similarity}.

    \item \textbf{Mapping-based methods}: These techniques transform the original data into representations that allow for efficient search. This category includes Locality-Sensitive Hashing (LSH) \cite{Indyk1998,Gionis1999,Datar2004,Andoni2006,Andoni2015} and Product Quantization approaches \cite{Jegou2011,Ge2013,Norouzi2013}, which encode vectors into compact codes that approximate the distances between points.

    \item \textbf{Navigable graph-based methods}: The most recent and currently state-of-the-art approach that constructs graph structures where vertices represent data points and edges connect similar points. These graphs are designed to allow ``greedy-like'' algorithms to perform a directed traversal through the data space.
\end{enumerate}

\subsection{Hierarchical Navigable Small World (HNSW)}

Hierarchical Navigable Small World (HNSW) \cite{hnsw} is a state of the art navigable graph-based method. It demonstrates superior performance on modern ANN benchmarks \cite{aumuller2020ann}. Due to relative simplicity, and efficiency, the HNSW algorithm has become widely used. It has been implemented in popular libraries for ANN-Search \cite{faiss}, \cite{nmspacelib}, and it is implemented as vector index in many database systems including PostgreSQL (vector extension), ORACLE \cite{OracleHNSW2023}, and other vector search oriented databases such as Milvus \cite{MilvusHNSW2024}, Zilliz \cite{ZillizHNSW2022}, Weaviate \cite{WeaviateVectorIndex2024}.  

\begin{figure}
  \centering
  \includegraphics[width=0.5\linewidth]{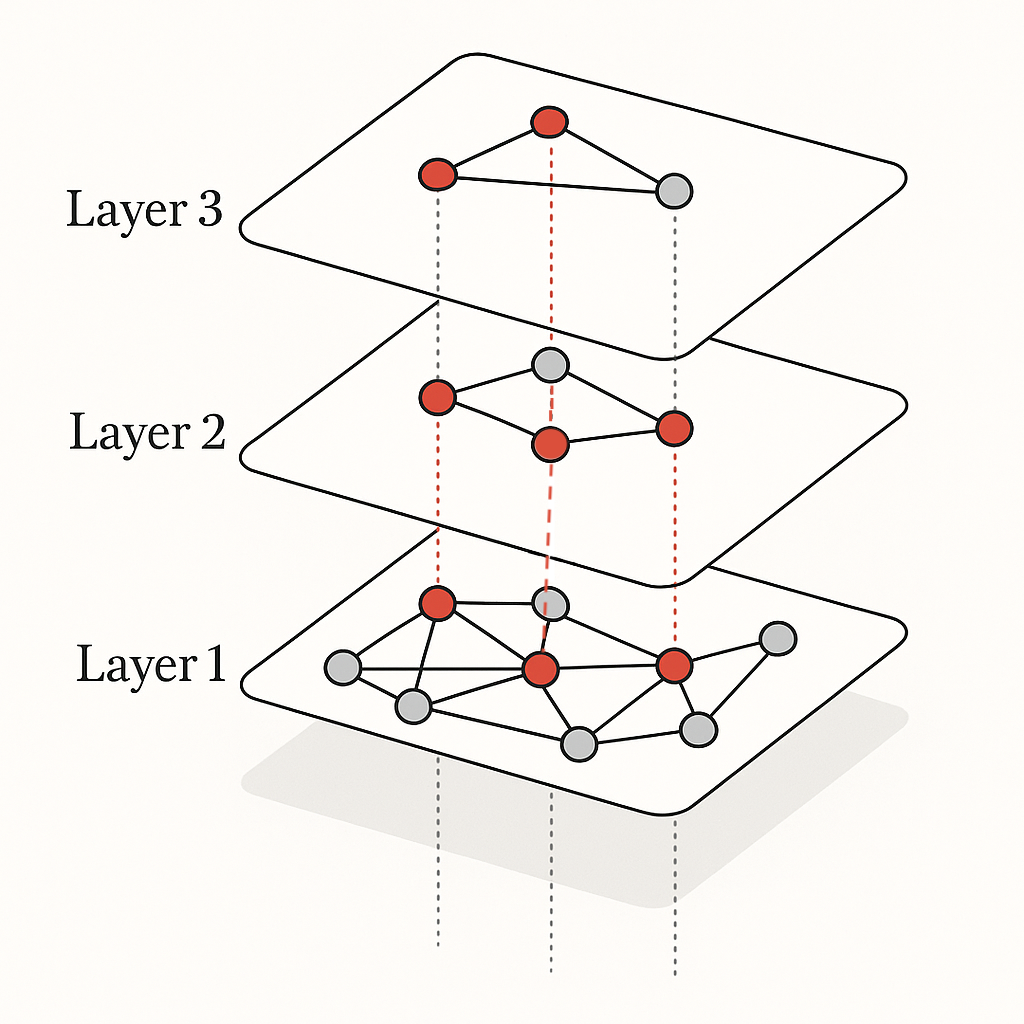}
  \caption{Schematic example HNSW structure. Each layer is a navigable graph that suit for ``greedy-like'' traversal algorithms. Graphs of all levels glued together form a navigable small world graph similar to NSW. An explicit separation set of edges of different lengths to layers helps to reduce the number of computations in search while traversing vertexes with high degrees, also called hubs. }
\label{fig:hnsw}
\end{figure}

HNSW \cite{hnsw} extends the navigable graphs concept by organizing points in a hierarchy of navigable graphs also named layers. Each layer contains a progressively smaller subset of data points. The bottommost layer contains all points, while the upper layers form a sparse representation of the data space. 
Graphs of all levels glued together form a navigable small world graph similar to NSW  \cite{nsw2011,nsw2012,nsw2014}. An explicit separation set of edges of different lengths to layers helps to reduce the number of computations in search while traversing vertexes with high degrees, also called hubs. A schematic example of HNSW structure presented in Fig. \ref{fig:hnsw}.

\subsection{Graph Merging Challenge}

While HNSW supports efficient dynamic insertions and deletions, an important operational challenge arises in practical systems: merging multiple independently constructed graph indices. This problem is particularly relevant in distributed systems, incremental indexing scenarios, and database compaction operations. In graph-based indices, deletion operations are often implemented by simply marking vertices as deleted without actually removing them, to avoid disrupting the graph's connectivity. This approach prevents the computationally expensive task of rebuilding neighborhoods for affected vertices. However, as deletions accumulate, the index becomes increasingly sparse and less efficient.

Graph merging provides a way to periodically compact the structure by creating a fresh graph that excludes deleted vertices while preserving the navigability properties essential for efficient search. Additionally, in distributed computing environments, merging separately constructed indices becomes necessary for building unified search structures over partitioned datasets.

\subsection{Graph Merging as an Iterative Process}

The merge operation can be conceptualized as an iterative process consisting of four key steps:

\begin{enumerate}
    \item \textbf{Processing Vertex Selection}: Choosing a vertex $v^*$ for which we aim to construct a neighborhood in the merged graph and determining which original graph it belongs to.
   
    \item \textbf{Candidate Collection}: Gathering potential neighbors from both input graphs. This involves searching within each graph structure to find vertices that should be considered for inclusion in $v^*$'s neighborhood.
   
    \item \textbf{Neighborhood Construction}: Applying a specified strategy (e.g., k-nearest neighbors or relative neighborhood graph) to select the final set of neighbors for $v^*$ from the candidates.
   
    \item \textbf{Information Propagation}: Deciding what information from the current iteration should be preserved for the next iteration to optimize the process, then proceeding to step 1 with a new vertex.
\end{enumerate}

By carefully designing each of these steps, we can create merge algorithms that balance computational efficiency with the quality of the resulting graph structure. 

In this work, we propose three novel algorithms based on this iterative framework—\textbf{Naive Graph Merge (NGM)}, \textbf{Intra Graph Traversal Merge (IGTM)}, and \textbf{Cross Graph Traversal Merge (CGTM)}—for efficiently merging HNSW graphs. These algorithms differ primarily in how they implement each step of the iterative process. For instance, NGM and IGTM select the next vertex for neighborhood construction only from the same graph as the current vertex. In contrast, CGTM implements the idea that the next vertex for neighborhood construction can be selected from any input graph.

The rest of this paper is organized as follows. Section 2 describes the search algorithms for HNSW graphs that we use in merge procedures. Section \ref{sec:neighborhood} recalls neighborhood construction strategies that are essential for both graph building and merging. Section \ref{sec:merge} introduces our three graph merging algorithms in detail. Section \ref{sec:experimens} presents computational experimental results. We give references to similar works in Section \ref{sec:relatedworks}. Finally, Section \ref{sec:conclusion} concludes the paper and discusses directions for future research.

\section{Search}
\label{sec:search}

The search process in navigable graphs is a critical operation that underlies both query processing and the graph construction phase. Below, we describe two key search algorithms: \textsc{LocalSearch} for exploring a single graph layer, and \textsc{HNSW-Search} for hierarchical multi-layer traversal.

\subsection{Local Search}

\begin{algorithm}
\caption{\textsc{LocalSearch}($G, q, C, k, L$)}\label{alg:local_search}
\textbf{Input:} Graph $G = (V, E)$, query $q \in \mathbb{R}^d$, initial candidate set $C \subset V$, $k \in \mathbb{N}$, $L \in \mathbb{N}$ \\
\textbf{Output:} Approximate $k$-nearest neighbors $V^* \subset V$
\begin{algorithmic}[1]
\While{True}
    \State $u \gets$ nearest unvisited point to $q$ in $C$
    \State $U \gets \{v \mid (u, v) \in E\}$
    \For{$v \in U$}
        \If{$v$ is unvisited}
            \State $C \gets C \cup \{v\}$
        \EndIf
    \EndFor
    \If{$|C| > L$}
        \State $C \gets$ top $L$ nearest points to $q$ in $C$
    \EndIf
    \If{no updates to $C$}
        \State \textbf{break}
    \EndIf
\EndWhile
\State \Return top-$k$ nearest points to $q$ in $C$
\end{algorithmic}
\end{algorithm}

The \textsc{LocalSearch} algorithm (Algorithm~\ref{alg:local_search}) implements a ``greedy'' like search within a single graph layer. Starting with an initial candidate set $C$, it iteratively explores the neighborhood of the closest unvisited point to the query $q$. This exploration strategy balances depth-first search for quick convergence toward the target region with breadth-first search for escaping local minima. In the literature, this algorithm is also named "beam search" \cite{prokhorenkova2020graph}, \cite{yang2024revisiting}. It was introduced in \cite{nsw2011} paper, and later adopted in HNSW in other graph-based methods as a basic search algorithm \cite{diskann}, \cite{wang2021comprehensive}.

For each iteration, \textsc{LocalSearch} selects the closest unvisited point $u$ to the query and explores its immediate neighbors. These neighbors are added to the candidate set if they have not been visited previously. The candidate set is constrained to size $L$ (the expansion factor) by retaining only the $L$ points closest to the query. This pruning step is crucial for an efficient search by focusing computational resources on the most promising candidates.

The search terminates when no additional updates to the candidate set occur during an iteration, indicating convergence. The algorithm then returns the $k$ nearest points to the query from the final candidate set, which represents the approximate $k$-nearest neighbors of the query within the graph structure.

The parameter $L$ provides a direct trade-off between search quality and computational cost. Larger values of $L$ allow for a broader exploration, potentially escaping local minima and finding better global solutions, but at the expense of increased computation time. In the HNSW paper, the parameter $L$ is named as \textbf{ef} (expansion factor) parameter.

\subsection{Hierarchical Search}
\label{subasec: hnsw-search}

\begin{algorithm}
\caption{\textsc{HNSW-Search}($\mathcal{H}, q, v_0, k, L, \ell$)}\label{alg:hnsw_search}
\textbf{Input:} HNSW graph $\mathcal{H} = (G_i)_{i=0}^{l_{\max}}$, query $q \in \mathbb{R}^d$, starting vertex $v_0 \in V$, $k, L \in \mathbb{N}$, search layer $\ell$ \\
\textbf{Output:} Approximate $k$-nearest neighbors $V^* \subset V$
\begin{algorithmic}[1]
\State $v^* \gets v_0$ 
\For{$i = l_{\max} \textbf{ down to } \ell$}
    \State $v^* \gets \textsc{LocalSearch}(G=G_i, q=q, C=\{v^*\}, k=1, L=L)$
\EndFor
\State \Return \textsc{LocalSearch}$(G_\ell, q, \{v^*\}, k, L)$
\end{algorithmic}
\end{algorithm}

The \textsc{HNSW-Search} algorithm (Algorithm~\ref{alg:hnsw_search}) leverages the hierarchical structure of HNSW to efficiently navigate through the vector space. The search begins at the highest layer $l_{\max}$ of the HNSW structure with a single entry point $v_0$. At each layer, \textsc{LocalSearch} is used to find the best approximation of the nearest neighbor to the query within that layer.

The algorithm traverses from the highest layer down to the target layer $\ell$, using the result from each layer as the entry point for the next lower layer. This coarse-to-fine approach allows the search to quickly focus on the relevant region of the vector space at higher, sparser layers before refining the search in the more densely connected lower layers.

At each layer $i$ (from $l_{\max}$ down to $\ell+1$), \textsc{LocalSearch} is executed with $k=1$ to find a single nearest neighbor to the query. This single neighbor serves as an entry point for the subsequent layer. At the target layer $\ell$, a final \textsc{LocalSearch} is performed with the specified value $k$ to retrieve the $k$-nearest neighbors.

This hierarchical search significantly reduces the number of distance computations compared to a flat graph search, especially for large-scale datasets. The efficiency comes from the progressive narrowing of the search space as the algorithm moves down through the layers, focusing the search on increasingly relevant regions.

Both search algorithms are fundamental not only for answering queries but also for the merge operations described in subsequent sections, as they provide the mechanism for identifying candidate neighbors when reconstructing connections in the merged graph.

\textsc{HNSW-Search} algorithm is a default search algorithm that is exposed to the user in all popular HNSW implementations with a default value of the parameter $\ell = 0$, and assuming that $v_0$ is one of the vertex of the top layer \cite{Vardanian_USearch_2023}, \cite{hnswlib2025}, \cite{faiss}.

In the text, we also refer to the \textsc{HNSW-Search} algorithm as a ``standard'', or ``default search" algorithm. The \textsc{LocalSearch} function is hidden from the user and is used only as a part of a particular implementation".

\section{Neighborhood Construction Strategies}
\label{sec:neighborhood}

\begin{algorithm}
\caption{\textsc{KNN-Neighborhood-Construction}($v^*, C, k$)}\label{alg:knntrategy}
\textbf{Input:} Vertex $v^*$, candidate set $C$, number of neighbors $k$ \\
\textbf{Output:} Set $C'$ of $k$ closest neighbors
\begin{algorithmic}[1]
\State $C' \gets$ $k$-nearest neighbors of $v^*$ in $C$
\State \Return $C'$
\end{algorithmic}
\end{algorithm}

Various strategies exist for selecting a node's neighborhood to form a navigable property for a graph. Our merging algorithms are designed to be agnostic to the specific neighborhood construction method, which can be provided as a parameter. We present two representative strategies below.

The \textsc{KNN-Neighborhood-Construction} algorithm (Algorithm~\ref{alg:knntrategy}) implements the simplest approach, where a vertex's neighborhood consists of its $k$-nearest neighbors from the candidate set. This strategy prioritizes proximity, but may not optimize for graph navigability properties.

In contrast, the \textsc{RNG-Neighborhood-Construction} algorithm (Algorithm~\ref{alg:rngstrategy}) implements a more sophisticated approach based on relative neighborhood graph principles. It processes candidates in ascending order of distance to the target vertex $v^*$. For each candidate $v$, it checks whether $v$ is closer to $v^*$ than to any previously selected neighbor $w$. This pruning condition helps create better-connected graphs with improved navigability by ensuring that edges span different directions in the vector space rather than clustering in the same region. The algorithm limits the neighborhood size to at most $m$ vertices to control maximum vertex degree.

The choice of neighborhood construction strategy significantly impacts both search performance and the computational cost of index maintenance operations, including merges. While \textsc{KNN-Neighborhood-Construction} is computationally simpler, \textsc{RNG-Neighborhood-Construction} typically produces graphs with better search performance at the expense of more complex neighborhood formation. In this paper, we performed our experiments with \textsc{RNG-Neighborhood-Construction}. 

\begin{algorithm}
\caption{\textsc{RNG-Neighborhood-Construction}($v^*, C, m$)}\label{alg:rngstrategy}
\textbf{Input:} Vertex $v^*$, candidate set $C$, maximum neighborhood size $m$ \\
\textbf{Output:} Filtered neighbor set $C'$
\begin{algorithmic}[1]
\State Sort $C$ in ascending order by distance to $v^*$
\State $C' \gets \emptyset$
\For{$v \in C$}
    \State $f \gets$ true
    \For{$w \in C'$}
        \If{$\rho(v^*, v) \geq \rho(v, w)$}
            \State $f \gets$ false
            \State \textbf{break}
        \EndIf
    \EndFor
    \If{$f$}
        \State $C' \gets C' \cup \{v\}$
    \EndIf
    \If{$|C'| \geq m$}
        \State \textbf{break}
    \EndIf
\EndFor
\State \Return $C'$
\end{algorithmic}
\end{algorithm}

\section{Merge Algorithms}
\label{sec:merge}

We now present three algorithms for merging HNSW graphs: \textsc{NGM}, \textsc{IGTM}, and \textsc{CGTM}. These algorithms operate layer by layer. We first describe the general procedure for merging the multi-layer HNSW structure, followed by the specific logic for merging a single layer (\textit{layer-merge algorithms}).

\subsection{Simple Insertion Graph Merge}
Before we start, we need to provide a short description of simple insertion graph merge (SIGM). It is a default merging strategy that uses an insertion operation to merge graphs. It takes the largest graph and sequentially inserts data from other graph. Actually, it is not a merging procedure, but we use it as a basic benchmark.

\subsection{HNSW General Merge Framework}

The \textsc{HNSW-General-Merge} algorithm (Algorithm~\ref{alg:general_merge}) provides a framework for merging two HNSW structures, $\mathcal{H}_a$ and $\mathcal{H}_b$. It iterates through each layer level and applies a chosen layer-merge algorithm (\textsc{NGM}, \textsc{IGTM}, or \textsc{CGTM}) to combine the corresponding layers $G^a_i$ and $G^b_i$. The layer-merge algorithms require access to the full HNSW structures $\mathcal{H}_a, \mathcal{H}_b$ to perform searches using \textsc{HNSW-Search}.

\begin{algorithm}
\caption{\textsc{HNSW-General-Merge}($\mathcal{H}_a, \mathcal{H}_b, \text{LayerMergeAlgo}, \text{params}$)}\label{alg:general_merge}
\textbf{Input:} HNSW graphs $\mathcal{H}_a$, $\mathcal{H}_b $. Chosen layer merge algorithm $\text{LayerMergeAlgo}$ (e.g., \textsc{NGM}). Algorithm-specific parameters $\text{params}$. \\
\textbf{Output:} Merged HNSW graph $\mathcal{H}_c = (G^c_i)_{i=0}^{l_{\max}^c}$
\begin{algorithmic}[1]
\State $l_{\max}^a \gets \mathcal{H}_a.\text{getMaxLayerNumber()}$
\State $l_{\max}^b \gets \mathcal{H}_b.\text{getMaxLayerNumber()}$
\State $l_{\max}^c \gets \max(l_{\max}^a, l_{\max}^b)$
\For{$i = 0 \textbf{ to } l_{\max}^c$} \Comment{Merge each layer}    
    \State $G^c_i \gets \text{LayerMergeAlgo}(\mathcal{H}_a, \mathcal{H}_b, \ell=i, \text{params})$ \Comment{Calls e.g., Merge-Naive}
\EndFor
\State $\mathcal{H}_c \gets (G^c_i)_{i=0}^{l_{\max}^c}$ 

\State \Return $\mathcal{H}_c$
\end{algorithmic}
\end{algorithm}



\subsection{Naive Graph Merge}
The \textsc{Naive Graph Merge} (\textsc{NGM}) algorithm (Algorithm~\ref{alg:merge_naive}) provides a straightforward method for merging a single layer $\ell$. The algorithm begins by extracting the target layers from both input HNSW structures and initializing the merged graph's vertex set as the union of vertices from both input graphs (lines 1-4).

For each vertex $v^*$ in graph $G^a_\ell$ (line 5), the algorithm:
\begin{enumerate}
  \item Searches for potential neighbors in graph $G^b_\ell$ using \textsc{HNSW-Search} (line 6)
  \item Combines these candidates with $v^*$'s original neighbors from $G^a_\ell$ (line 7)
  \item Selects the final neighborhood for $v^*$ using the specified \textsc{NeighborhoodConstruction} strategy and adds the corresponding edges to $E^c$ (line 8)
\end{enumerate}

The same process is then repeated for all vertices of graph $G^b_\ell$ (lines 9-12).

This approach ensures that each vertex in the merged graph has an appropriate neighborhood that incorporates information from both input graphs. However, it is computationally intensive due to repeated \textsc{HNSW-Search} calls that traverse multiple layers for each vertex.

\begin{algorithm}
\caption{\textsc{NGM}($\mathcal{H}_a, \mathcal{H}_b, \ell, \text{NeighborhoodConstruction}, m, \text{search\_ef}, v_{entry}$)}\label{alg:merge_naive}
\textbf{Input:} HNSW graphs $\mathcal{H}_a$, $\mathcal{H}_b$; target layer $\ell$; Neighborhood construction function \textsc{NeighborhoodConstruction}; target neighborhood size $m$; search parameter $\text{search\_ef}$; entry point $v_{entry}$ (e.g., from $\mathcal{H}_a$ or $\mathcal{H}_b$) \\
\textbf{Output:} Merged graph $G^c$
\begin{algorithmic}[1]
\State $G^a \gets \mathcal{H}_a\text{.GetLayer}(\ell) $; $G^b \gets\mathcal{H}_a\text{.GetLayer}(\ell)$ 
\State $V^a \gets \text{vertices}(G^a)$; $V^b \gets \text{vertices}(G^b)$
\State $E^a \gets \text{edges}(G^a)$; $E^b \gets \text{edges}(G^b)$
\State $V^c \gets V^a \cup V^b$; $E^c \gets \emptyset$ 

\For{$v^* \in V^a$} 
    \State $\mathcal{C}^b \gets \textsc{HNSW-Search}(\mathcal{H}=\mathcal{H}_b, q=v^*, v_{entry}=v_{entry}^b, k=m, L=\text{search\_ef}, \ell=\ell)$ 
    \State $\mathcal{C} \gets \{v \mid (v^*, v) \in E^a \} \cup \mathcal{C}^b$  
    \State $E^c \gets E^c \cup \{(v^*, v) | v \in \textsc{neighborhood\_construction}(\mathcal{C}, v^*, m)\}$
\EndFor

\For{$v^* \in V^b$} 
    \State $\mathcal{C}^a \gets \textsc{HNSW-Search}(\mathcal{H}=\mathcal{H}_a, q=v^*, v_{entry}=v_{entry}^a, k=m, L=\text{search\_ef}, \ell_{target}=\ell)$ 
    \State $\mathcal{C} \gets \{v \mid (v^*, v) \in E^b \} \cup \mathcal{C}^a$  
    \State $E^c \gets E^c \cup \{(v^*, v) | v \in \textsc{neighborhood\_construction}(\mathcal{C}, v^*, m)\}$
\EndFor

\State \Return $G^c = (V^c, E^c)$
\end{algorithmic}
\end{algorithm}

\subsection{Intra Graph Traversal Merge}

\begin{algorithm}
\caption{\textsc{IGTM}($\mathcal{H}_a, \mathcal{H}_b, \ell, \text{jump\_ef}, \text{local\_ef}, \text{next\_step\_k}, M, m$)}\label{alg:IGTM}
\textbf{Input:} The HNSW graphs $\mathcal{H}_a = (G^a_i), \mathcal{H}_b = (G^b_i)$, the merging layer number $\ell$, the size of the forming neighborhoods $m \in \mathbb{N}$, parameters $\text{jump\_ef}, \text{local\_ef}, \text{next\_step\_k} \in \mathbb{N}$ \\
\textbf{Output:}  Merged graph $G^c$ 
\begin{algorithmic}[1]
\State $G^a \gets \mathcal{H}_a\text{.GetLayer}(\ell) $; $G^b \gets\mathcal{H}_a\text{.GetLayer}(\ell)$ 
\State $V^a \gets \text{vertices}(G^a)$; $V^b \gets \text{vertices}(G^b)$
\State $E^a \gets \text{edges}(G^a)$; $E^b \gets \text{edges}(G^b)$
\State $V^c \gets V^a \cup V^b$; $E^c \gets \emptyset$ 
\State $\mathcal{V}_{not\_done} \gets V^a$

\While{$\mathcal{V}_{not\_done} \neq \emptyset$}
    \State $v^* \gets \text{random choice from } \mathcal{V}_{not\_done}$
    
    \State $\mathcal{P}^b  \gets \textsc{HNSW-Search}(\mathcal{H}=\mathcal{H}^b, q=v^*, v_0, k=M, L=\text{jump\_ef}, \ell)$
    
    \While{True}
        \State $\mathcal{V}_{not\_done} \gets \mathcal{V}_{not\_done} \setminus \{v^*\}$
        
        \State $\mathcal{C}^b  \gets \textsc{LocalSearch}(G=G^b, q=v^*, C=\mathcal{P}^b , k=m, L=\text{local\_ef})$
        
        \State $\mathcal{C} \gets  \{v : (v^*, v) \in E^a \} \cup \mathcal{C}^b$
        
        \State $E^c \gets E^c \cup  \{ (v^*, v)  : v \in \textsc{NeighborhoodConstruction}(\mathcal{C}, v^*, m) \}$

        \State $\mathcal{P}^b \gets \{\mathcal{C}^b_1, \mathcal{C}^b_2, ..., \mathcal{C}^b_M \} $

        \State $\mathcal{C}^a  \gets \textsc{LocalSearch}(G=G^a, q=v^*, C=\{v^*\} , k=\text{next\_step\_k}, L=\text{next\_step\_ef})$
        
        \State $\mathcal{C}^a \gets \mathcal{C}^a \cap \mathcal{V}_{not\_done}$
        
        \If{$\mathcal{C}^a = \emptyset$}
            \State \textbf{break}
        \EndIf
        
        \State $v^* \gets \mathcal{C}^a_1$
    \EndWhile
\EndWhile
\State $\mathcal{V}_{not\_done} \gets V^b$
\While{$\mathcal{V}_{not\_done} \neq \emptyset$}
    \State Repeat the same process for $V^b$ with the roles of $\mathcal{H}_a$ and $\mathcal{H}_b$ swapped.
\EndWhile
\State \Return $G^c=(V^c,E^c)$
\end{algorithmic}
\end{algorithm}

The most effort of the \textsc{NGM} algorithm lies in obtaining the set of neighborhood candidates from the other graph utilizing the \textsc{HNSW-Search} procedure, which every time traverses the layer graphs from the top level down to the layer number $\ell$. The number of computations can be reduced if we select the next vertex to process $v^*$ close to the previous one (line 15 of the algorithm ~\ref{alg:IGTM}), instead of randomly choosing it. Thus, for the new $v^*$ the neighborhood candidates will also be close to the previous candidates set. To search for these new neighborhood candidates we can use the \textsc{LocalSearch} procedure which traverses the same graph staring from the previous neighborhood candidates set $\mathcal{P}^b$ (lines 11). In line 14 in the set $\mathcal{P}^b$ we keep only $M$-closest to $v^*$ candidates.\\
In lines 15,16, we select a new processing vertex $v^*$  close to the previous processing vertex $v^*$ that was not already processed.  
To ensure that the new processing vertex $v^*$ is not very far from the previous $v^*$ we bound the size of the results of the \textsc{LocalSearch} procedure controlling by next\_step\_k parameter. Once \textsc{LocalSearch} cannot find enough close unprocessed vertex (if condition in line 17), in line 7 we choose a new $v^*$ from the set $\mathcal{V}_{not\_done}$ randomly.\\
After we have processed all vertices from the graph $G^a$. We do the same for the vertices of graph $G^b$ (lines 21-22).  

\subsection{Cross Graph Traversal Merge}

Cross Graph Traversal Merge (\textsc{CGTM}) algorithm is similar to \textsc{IGTM} utilizes the \textsc{LocalSearch} procedure to reduce computation efforts. The difference is that the \textsc{IGTM} algorithm chooses the next processing vertex $v^*$ from the same graph, while \textsc{CGTM} looks for the new processing vertex $v^*$ in both graphs $G^a$, and $G^b$. 
Thus in line 24 $v^*$ is chosen from the set $\mathcal{C}_{not\_done}$, which is one of the not processed vertex of both graph (line 21). The intuition underlying \textsc{CGTM} algorithm is that when we choose a new processing vertex $v^*$ from both graphs, we reduce the number of times when $v^*$ is chosen randomly, thus minimizing the number of times that we use the more expensive search procedure \textsc{HNSW-Search}.

\begin{algorithm}
    \caption{\textsc{CGTM}($\mathcal{H}_a, \mathcal{H}_b, \ell, \text{jump\_ef}, \text{local\_ef}, \text{next\_step\_k}, M, m$)}\label{alg:CGTM}
    \textbf{Input:} The HNSW graphs $\mathcal{H}_a = (G^a_i)$, $\mathcal{H}_b = (G^b_i)$, the merging layer $\ell$, parameters $\text{jump\_ef}, \text{local\_ef}, \text{next\_step\_k} \in \mathbb{N}$, neighborhood sizes $M, m \in \mathbb{N}$ \\
    \textbf{Output:} Merged graph $G^c$ 
\begin{algorithmic}[1]

\State $G^a \gets \mathcal{H}_a\text{.GetLayer}(\ell) $; $G^b \gets\mathcal{H}_a\text{.GetLayer}(\ell)$ 
\State $V^a \gets \text{vertices}(G^a)$; $V^b \gets \text{vertices}(G^b)$
\State $E^a \gets \text{edges}(G^a)$; $E^b \gets \text{edges}(G^b)$
\State $V^c \gets V^a \cup V^b$; $E^c \gets \emptyset$
\State $\mathcal{V}_{not\_done} \gets V^a \cup V^b$

\While{$\mathcal{V}_{not\_done} \neq \emptyset$}
    \State $v^* \gets \text{randomly choice from } \mathcal{V}_{not\_done}$

    \State $\mathcal{P}^a  \gets \textsc{HNSW-Search}(\mathcal{H}=\mathcal{H}^a, q=v^*, v_0, k, L=\text{jump\_ef}, \ell $)

    \State $\mathcal{P}^b \gets \textsc{HNSW-Search}(\mathcal{H}=\mathcal{H}^b, q=v^*, v_0, k, L=\text{jump\_ef}, \ell $)

    \While{True}
        \State $\mathcal{V}_{not\_done} \gets \mathcal{V}_{not\_done} \setminus \{v^*\}$

        \State $ \mathcal{C}^a  \gets \textsc{LocalSearch}(G=G^a, q=v^*, C=\mathcal{P}^a  , k=m, L=\text{local\_ef})$
        
        \State $\mathcal{C}^b \gets \textsc{LocalSearch}(G=G^b, q=v^*, C=\mathcal{P}^b , k=m, L=\text{local\_ef})$
        
        \If{$v^* \in V^a $}
            \State $\mathcal{C}\gets  \{v : (v^*, v) \in E^a \} \cup  \mathcal{C}^b\}$
        \Else
            \State $\mathcal{C} \gets  \{v : (v^*, v) \in E^b \} \cup  \mathcal{C}^a\}$
        \EndIf
        
        \State $E^c \gets E^c \cup \{ (v^*, v) : v \in \text{neighborhood\_construction}(\mathcal{C}, v^*, m) \}$
        
        \State $\mathcal{C}^a_{not\_done} \gets \{\mathcal{C}^a_1, \mathcal{C}^a_2, ..., \mathcal{C}^a_{ \text{next\_step\_k} } \} \cap \mathcal{V}_{not\_done}$

        \State $\mathcal{C}^b_{not\_done} \gets \{\mathcal{C}^b_1, \mathcal{C}^b_2, ..., \mathcal{C}^b_{ \text{next\_step\_k} } \}  \cap \mathcal{V}_{not\_done}$

        \State $\mathcal{C}_{not\_done} \gets \mathcal{C}^a_{not\_done} \cup \mathcal{C}^b_{not\_done}$
        
        \If{$\mathcal{C}_{not\_done} = \emptyset$}
            \State \textbf{break}
        \EndIf
        
        \State $v^* \gets \underset{v \in \mathcal{C}_{not\_done}}{\mathrm{argmin}} \rho(v, v^*)$
        
        \State $\mathcal{P}_a \gets \mathcal{C}^a$
        \State $\mathcal{P}_b \gets \mathcal{C}^b$
    \EndWhile
\EndWhile
\State \Return $G^c=(V^c,E^c)$
\end{algorithmic}
\end{algorithm}

\section{Computational Experiments}
\label{sec:experimens}

\begin{figure}
  \centering
  \includegraphics[width=1.\linewidth]{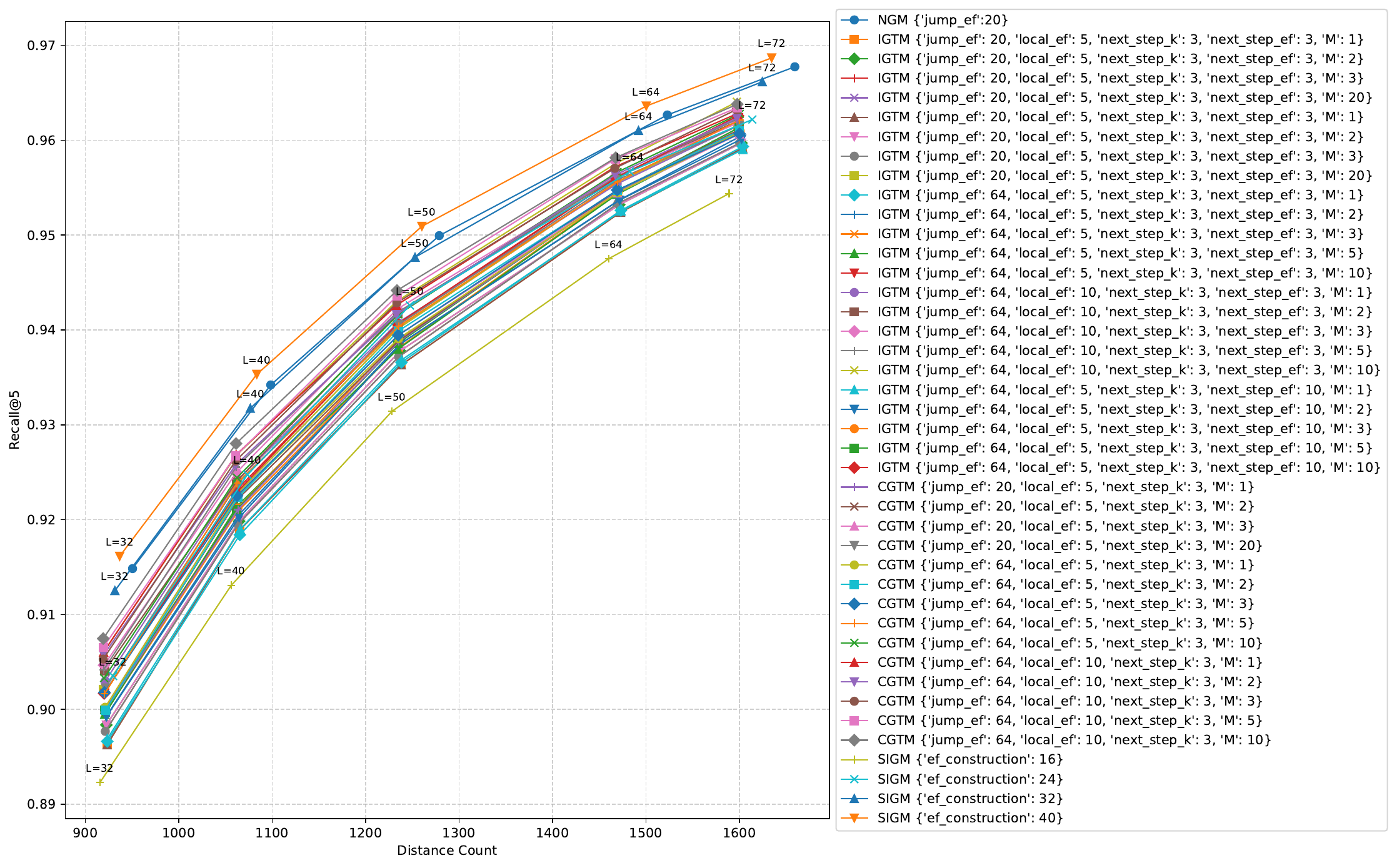}
  \caption{Recall vs distance count on search stage for merged graphs. }
\label{fig:search}
\end{figure}




\subsection{Setup}

We evaluated the proposed merge algorithms on the standard ANN benchmark dataset SIFT1M of 1 million 128-dimensional vectors. We divided it into two disjoint subsets of 500k vectors. In these subsets, we built two HNSW indices $\mathcal{H}_a$ and $\mathcal{H}_b$ using the following parameters: $M = 16$, $M0$ = 32, ef\_construction = 32.
Then these indices were merged using a simple insertion strategy \textsc{SIGM} and the three proposed algorithms: \textsc{NGM}, \textsc{IGTM}, and \textsc{CGTM}. We keep the same values of the parameters $m = 16$ and $m0 = 32$ as the target neighborhood size in the merged graph.

In order to be independent of implementation details and low-level optimisations, we made a comparison of the algorithms based on the number of distance computations required for the merge process.

Other important characteristics of the merge process is how accurate the merge is done. In this case, the word ``merging accuracy'' can mean different things. For example, we can estimate merging accuracy as how similar the neighborhoods of the merged graph to the neighborhoods of the graph constructed by simple insertion strategy \textsc{SIGM}. 
Another way, which we prefer, is to verify how the merged graph is good for search. To quantify search quality, we performed a standard search performance test on the merged graphs. So, for the merged graphs, we have measured a trade-off between recall and the number of distance computations, by running algorithm ~\ref{alg:hnsw_search} with different values of the parameter $L$ (search expansion factor). We measured the recall@5 metric for L = 32, 40, 50, 64, and 72. This metric indicates how often the true nearest neighbors appear in the top-5 results returned by the search algorithm.

Formally, for a given query point \(q\), let \(P_k(q)\) denote the set of \(k\) true nearest neighbors, and let \(A_k(q)\) be the set of the top \(k\) points returned by the algorithm. Then, recall@k is defined as:

\[
\text{recall@}k = \frac{|P_k(q) \cap A_k(q)|}{k}
\]
In our experiment as recall@5 we report an averaged value over all sequences of searches.

All source codes of our implementation can be found in a public GitHub repository \footnote{\url{https://github.com/aponom84/merging-navigable-graphs}}.

\subsection{Results}

The searching quality of the merged graph is represented in Fig. ~\ref{fig:search}. The graph plots trade-off between recall and the number of distance computations. As can be seen, \textsc{SIGM} with ef\_construction $\ge$ 32 and \textsc{NGM} with parameter jump\_ef = 20 produce merged graph with slightly better searing quality than other algorithms.

\begin{figure}
  \centering
  \includegraphics[width=1.\linewidth]{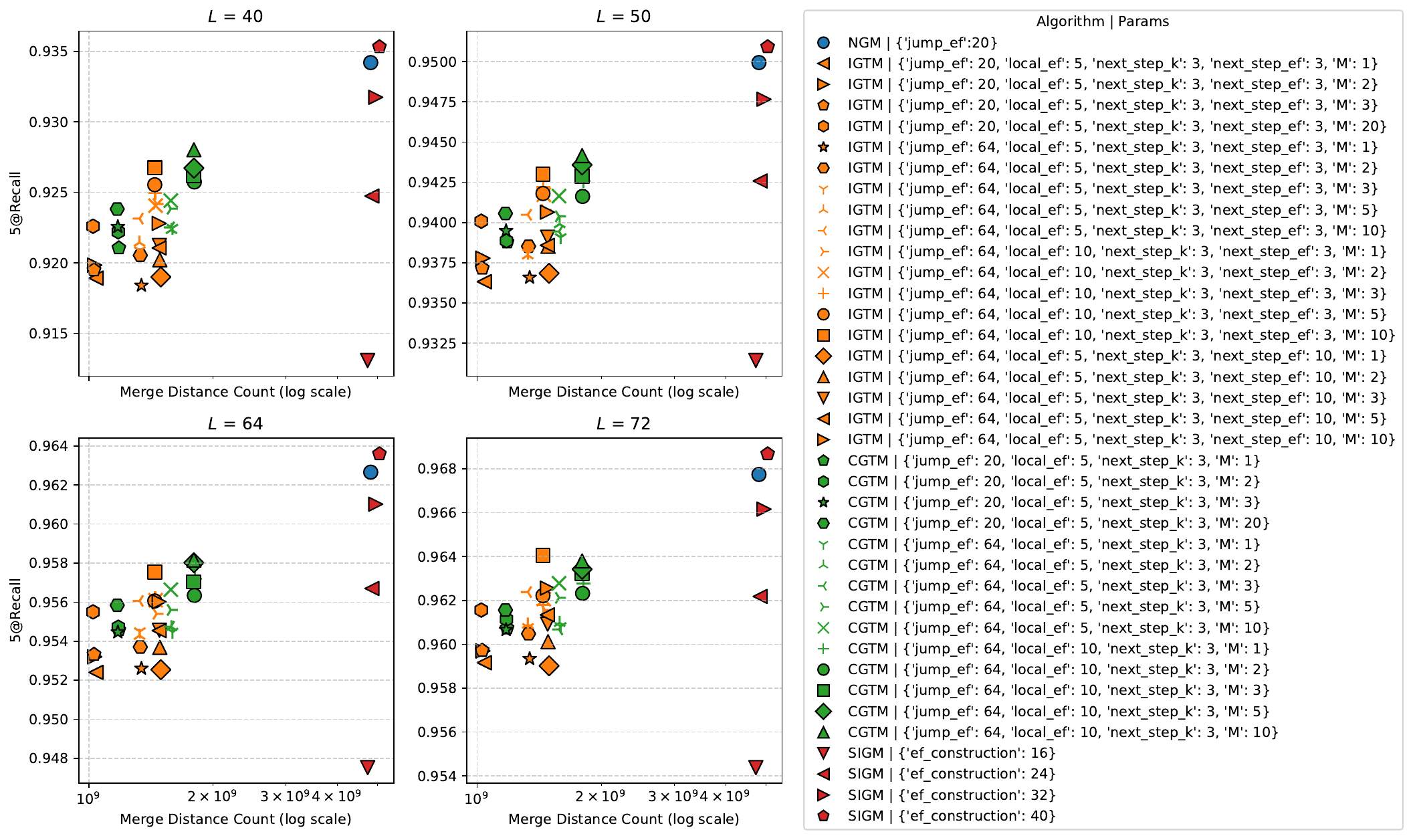}
  \caption{Recall of final merged graph vs merging efforts}
  \label{fig:recall_vs_merge}
\end{figure}

Another important fact which ~\ref{fig:search} shows that the number of distance computations performed by the search algorithm (Algorithm ~\ref{alg:hnsw_search}) for equal parameter $L$ are almost the same.  Therefore, the graph is better if it provides better search recall for a fixed value of the parameter $L$. Taking this into account, we can better interpret Fig. ~\ref{fig:recall_vs_merge}. It is easy to see the following:  

\begin{itemize}
    \item\textsc{IGTM} and \textsc{CGTM} in our experiment setup  achieve recall better than \textsc{SIGM} with parameter ef\_construciton = 24;
    \item \textsc{NGM} and \textsc{SIGM} perform the most computations, but achieves the highest recall, as it exhaustively reconstructs each neighborhood;
    \item \textsc{CGTM} achieving comparable recall with approximately 60\% fewer computations than \textsc{NGM} and \textsc{SIGM};
    \item \textsc{IGTM} provides the best run-time performance, reducing distance computations about 20\% than \textsc{CGTM}, and about 70\% than \textsc{NGM} and \textsc{SIGM} while achieving comparable recall.
\end{itemize}

\section{Related Works on Graph Merging}
\label{sec:relatedworks}
While working on the paper, we have found that just recently researchers from \textbf{Elastic Search Lab} published a blog post \cite{ElasticHNSW2023} where they proposed a fast merge algorithm, which also utilizes information about closeness of object in all graphs. The authors presented a scheme where a subset of vertices $J$ ``join set'' from the smaller graph is selected by greedy heuristic and inserted into the larger graph using standard HNSW insertion procedure. For the remaining vertices from the small graph, their already inserted neighbors and those neighbors' connections in the large graph are used to perform a limited beam search (named FAST-SEARCH-LAYER) to find connection candidates. This approach limits full insertions and accelerates the merging process.

Ideologically approach of \cite{ElasticHNSW2023} is similar to our work, because the authors utilize information about closeness of object in all graphs, and use light-weighted version of the search algorithm FAST-SEARCH-LAYER (in our case it is \textsc{LOCAL SEARCH}. The ``join set'' $J$ from some point of view is similar to the set vertices for which \textsc{IGTM} and \textsc{CGTM} run the standard HNSW search. However, in \cite{ElasticHNSW2023} the ``join set'' $J$ is determined separately at the first stage by a greedy heuristic that tries to find a good cover set, while \textsc{IGTM} and \textsc{CGTM} establish the vertices for which to run \textsc{HNSW-Search} during the work in the main cycle in a more homogeneous way. 

The other work which is relatively close to the present topic is a paper by Zhao and co-authors \cite{knnmerge}. They proposed two algorithms: Symmetric Merge (S-Merge) for combining two k-NN graphs and Joint Merge (J-Merge) for incrementally extending an existing graph with new points. These methods effectively combine graphs built on different subsamples while maintaining high search accuracy and accelerating computations. S-Merge discards half the neighbors in each list, while J-Merge leverages the hierarchical structure of the graph.

The performance comparison of the different merge algorithms is a subject of future research. 

\section{Conclusion and Future Work}
\label{sec:conclusion}

In this work, we introduced three algorithms to merge hierarchical navigable small-world graphs: \textsc{NGM} (Algorithm~\ref{alg:merge_naive}), \textsc{IGTM} (Algorithm~\ref{alg:IGTM}), and \textsc{CGTM} (Algorithm~\ref{alg:CGTM}). 

The \textsc{NGM} algorithm provides a straightforward but computationally intensive approach, performing standard HNSW searches to find sets of candidates to reconstruct the neighborhood of a vertex chosen in an arbitrary way. \textsc{IGTM} and \textsc{CGTM} improve efficiency by leveraging locality—processing vertices close to each other sequentially and using the less expensive \textsc{LocalSearch} algorithm (Algorithm~\ref{alg:local_search}).

Our experimental results demonstrate that \textsc{IGTM} and \textsc{CGTM} significantly reduce the number of distance computations compared to the naive approach while maintaining comparable search accuracy. Our evaluation on the dataset of 1 million 128-dimensional vectors (SIFT1M) shows that \textsc{IGTM} and \textsc{CGTM} are more than 3 times faster than the straightforward insertion strategy \textsc{SIGM}, with minimal impact on recall performance.

Interestingly, our experiments revealed that \textsc{IGTM} outperformed \textsc{CGTM} in terms of computational efficiency, contrary to our initial expectations. We had anticipated that \textsc{CGTM} would be more efficient since it can select the next vertex for neighborhood construction from both graphs, theoretically reducing the chances of getting stuck and thus requiring fewer costly standard search operations. However, it appears that the overhead of selecting the next processing vertex, for which a neighborhood is forming, in \textsc{CGTM} is too computationally expensive, and these costs are not offset by the reduction in searches when ``gets stuck'' with selecting the next close vertex to process. The study of this fact is a subject for further work.

In addition, an important direction for future work is adapting the proposed merge algorithms to handle deleted vertices. This would enable their use in compaction processes, where graphs are periodically restructured to remove obsolete entries and maintain search efficiency. By extending our merge algorithms to filter out deleted nodes during the ``Processing Vertex Selection'' phase.

Also, the researchers can focus on the idea that the neighborhood construction can be done for the set of vertex close to each other instead of forming neighborhood for one vertex per iteration. This can be more suitable for GPU or NPU settings.  

Additional areas for exploration include adaptive parameter selection based on dataset characteristics, and extensions to other graph-based index structures beyond HNSW. 

Finally, future work can be applied to develop a fast graph construction procedure using one of the merge algorithms. It seems very natural to start from the set of small graphs and merge them in a recursive manner.

\bibliographystyle{unsrt}

\end{document}